# LGBTQIA+ (In)Visibility in Computer Science and Software Engineering Education

Ronnie de Souza Santos
Cape Breton University
Sydney, NS, Canada
ronnie_desouza@cbu.ca

Brody Stuart-Verner
Cape Breton University
Sydney, NS, Canada
brody_verner@cbu.ca

Cleyton V. C. de Magalhaes
CESAR School
Recife, PE, Brazil
cvcm@cesar.school

*Abstract*— **Modern society is diverse, multicultural, and multifaceted. Because of these characteristics, we are currently observing an increase in the debates about equity, diversity, and inclusion in different areas, especially because several groups of individuals are underrepresented in many environments. In computer science and software engineering, it seems counter-intuitive that these areas, which are responsible for creating technological solutions and systems for billions of users around the world, do not reflect the diversity of the society to which it serves. In trying to solve this diversity crisis in the software industry, researchers started to investigate strategies that can be applied to increase diversity and improve inclusion in academia and the software industry. However, the lack of diversity in computer science and related courses, including software engineering, is still a problem, in particular when some specific groups are considered. LGBTQIA+ students, for instance, face several challenges to fit into technology courses, even though most students in universities right now belong to Generation Z, which is described as open-minded to aspects of gender and sexuality. In this study, we aimed to discuss the state-of-art of publications about the inclusion of LGBTQIA+ students in computer science education. Using a mapping study, we identified eight studies published in the past six years that focused on this public. We present strategies developed to adapt curricula and lectures to be more inclusive to LGBTQIA+ students and discuss challenges and opportunities for future research.**

*Keywords—EDI, LGBTQIA+, software engineering, inclusion, diversity, education*

## I. Introduction

Even though equity, diversity, and inclusion (EDI) are currently under discussion in different contexts of modern society, recent studies continuously highlight how our society is underrepresented in some areas. In some specific fields, such as Science, Technology, and Engineering, the levels of EDI are particularly low [1]. This reality is at least curious because modern society strongly relies on technology (e.g., software) for all sorts of processes and activities. Hence, why some groups of individuals are excluded from the process of conceiving such technologies?

Looking at the context of computer science (CS) and related fields, e.g., Software Engineering (SE), recent research identified and analyzed a series of studies published over the years discussing diversity in the software industry. The authors highlighted the lack of representativeness in software companies and discussed how diversity is essential for software teams, as a variety of backgrounds could support teams in producing better results [2].

The lack of diversity in the software industry likely reflects the lack of diversity in academia. This hypothesis is supported by studies that demonstrated that participation in computer science departments frequently under-represents many groups [3][4][5][6], resulting in a reduction of the diversity in groups of newly graduated professionals. This unequal scenario is complex, as it involves the intersection of historical, structural, social, and cultural aspects [4][7]. Even though coping strategies have been proposed over the years, an egalitarian environment in the technology area is still far from satisfactory [8]. In particular, previous reviews indicate that among the under-represented groups, there are groups that have received even less attention than others [2][8], such as LGBTQIA+ individuals.

As the research on equality, diversity, and inclusion in technology intensifies, it is possible to notice an increase in the efforts of solving the diversity problem in the academic context (i.e., schools, colleges, and universities), possibly as an attempt to propagate this diversity into the software industry or because Generation Z, which is more diverse, is currently predominant in the universities' demographic [30] and will be most of the professionals arriving in the software industry soon.

In this sense, we aim to contribute to the characterization of the research on EDI in CS and SE education. Thus, we conducted a mapping study to identify and explore strategies and experiences regarding the inclusion of the LGBTQIA+ community in CS and SE courses. LGBTQIA+ is one of the under-represented groups that received little attention from academic research on technology education over the years [2][8], even though one of the most important names in the field belonged to this group, namely Alan Turing. Thus, the following research question guided this research:

*RQ. What is currently known about the inclusion of LGBTQIA+ students in computer science education?*

From this introduction, the present study is organized as follows. In Section 2, we present our theoretical background. Section 3 presents how we designed and conducted our mapping study. In section 4, we present our results. In section 5, such results are discussed, and the main implications of the study are highlighted. Finally, in section 6, we present our conclusions and directions for future research.

## II. EDI in Education

EDI is a complex phenomenon centered on developing approaches for providing equal opportunity for individuals (equity) while recognizing their personal, social, and cultural differences (diversity) and encouraging them to participate in spaces and debates (inclusion) [9]. Modern society is diverse, multicultural, and multifaceted [11]; this is why we have recently observed an increase in EDI debates.

These debates have arrived at the universities [10], where discussions about inclusion are changing the scenario of courses known for being mostly homogeneous, i.e., with little or no diversity. In this context, academics started to develop and implement teaching strategies, activities, and practices that could be adapted across disciplines to foster equity, diversity, and inclusion in classrooms [12].

In health courses, educators focus on diversity to promote a healthcare workforce that represents the various aspects of communities regarding gender, ethnicity, sexual orientation, and other backgrounds [14] [13]. In fields related to business and management, diversity and inclusion in undergraduate programs have been studied for over a decade since EDI plays a vital role in organizations [15][16]. In science, technology, engineering, and mathematics, few experiences are reported regarding increasing diversity and inclusion in undergraduate programs; however, it is possible to observe a particular increase in the studies focusing on gender inclusion [17][18].

EDI discussions are relatively new in software engineering [19], even though the lack of diversity in the software industry is causing several problems for individuals, including the development of computational algorithms that discriminate against non-white communities and restrict groups of individuals from services [20]; hostile and unfair software development environments, especially for women [21][22]; and non-inclusive academic environments [23].

A general agreement among these knowledge areas is that universities are one of the most significant institutions to promote diversity and inclusion because if higher education is transformative enough to former professionals that recognize the importance of having diverse and inclusive environments, we will likely have workforce leaders who have the disposition, knowledge, and skills to propagate such transformation across industries and the society [10].

III. METHOD

Secondary studies are a type of research commonly applied to integrate results obtained from several primary studies (such as experiments, case studies, surveys, and experience reports, among others) [10]. The two most common types of secondary studies conducted in software engineering are the systematic literature review and the mapping study (or scoping review). Conventional systematic reviews aggregate evidence about a specific problem and are applied to solve relational and comparative research questions. On the other hand, mapping studies are a particular type of systematic review with a broader view of primary studies, focused on answering descriptive questions and revealing trending topics in the research field [25].

This study can be classified as a mapping study. A mapping study frequently answers questions related to trends in research, e.g., what do we know about topic T? [10]. Following this definition, the present study aims to identify papers and articles published over the years that simultaneously addressed topics related to computer science education (which includes software engineering courses) and strategies to foster the inclusion of LGBTQIA+ students. The main goal is to construct an overview of the theme that can guide the next steps of our research focused on diversity in CS and SE programs and identify opportunities for future research. To achieve this goal, we followed the guidelines for performing systematic reviews in software engineering [26], as presented below.

*A. Research Questions*

The primary goal of this study is to draw the panorama of the research on the inclusion of LGBTQIA+ students in CS and SE education. Based on the primary research goal of this study, we defined four specific research questions to guide and structure the extraction, analysis, and synthesis of evidence obtained from papers and articles published over the years:

- RQ1. What is the evolution in the number of studies over the years?

- RQ2. What experiences have been reported on including LGBTQIA+ students in CS and SE courses?

- RQ3. What are the future challenges regarding increasing diversity and inclusion of LGBTQIA+ students in CS and SE courses?

- RQ4. What are the implications of this panorama for the software industry and research community?

*B. Data Sources and Search Strategy*

We started the search process with an automated strategy by applying a search string on five search engines and indexing systems (Table 1). The search string was based on terms extracted from the general RQ (Figure 1). We kept the string based only on the term *computer science* because a) undergraduate courses in computer science often offer several courses on software engineering, and b) specific conferences in the context of software engineering were addressed manually posteriorly. The automated search process performed in early 2022 retrieved 194 papers presenting findings about EDI in the context of computer science education.

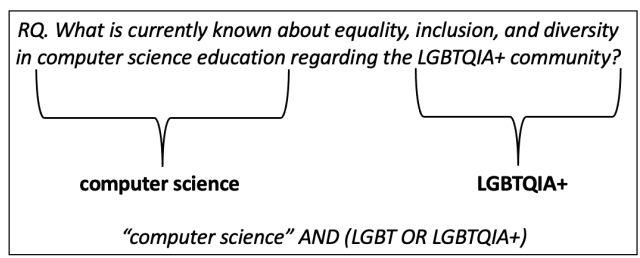

Fig. 1. Search string.

TABLE I. SOURCES FOR AUTOMATIC SEARCH

| Search Engine | Link |
|---|---|
| ACM Digital Library | http://dl.acm.org/ |
| IEEEXplore | https://ieeexplore.ieee.org/Xplore/home.jsp |
| Scopus | https://www.scopus.com/ |
| Springer | https://link.springer.com/ |
| Google Scholar | https://scholar.google.com.br/ |

*For Google Scholar, the first five pages of results were considered.

Following this, we conducted a manual search to manually checked the proceedings of four conferences related to computer science and software engineering education and one conference focused on EDI in computer science. We conducted the manual search in an amount of 1416 papers published in the following conferences:

- 2017 – 2022: The Technical Symposium on Computer Science Education (SIGCSE), the Conference on International Computing Education Research (ICER), and the International Conference on Software Engineering – Software Engineering Education and Training (ICSE - SEET).

- 2018 – 2022: International Conference on Research in Equity and Sustained Participation in Engineering, Computing, and Technology (RESPECT).

- 2021 and 2022: International Workshop on Software Engineering Education for the Next Generation (SEENG)

The manual search was planned to cover at least five years of all selected conferences above. However, for RESPECT and SEENG, the proceedings of some editions were not available online. The manual search process was performed in mid-2022, and 73 papers about EDI in education were identified.

*C. Inclusion and Exclusion Criteria*

The initial set of 267 papers addressing aspects of EDI in CS and SE education retrieved from the combination of automated and manual searches was submitted to two filtering cycles.

First, papers were excluded whenever they met any of the following exclusion criteria:

(1) Written in any language but English.

(2) Not available to download.

(3) Incomplete papers, drafts, and presentation slides.

(4) Papers that only cite LGBTQIA+ students (e.g., in a definition, contextualization, or background section).

(5) Papers that do not present findings or discussions on LGBTQIA+ in CS and SE education.

Second, we selected studies presenting findings, discussions, lessons learned, and/or experience reports discussing educational topics that addressed LGBTQIA+ students in CS and SE courses.

*D. Studies Selection*

Initially, papers were selected based on the analysis of metadata (e.g., title, abstract, references, and keywords) following the five exclusion criteria. Then, the inclusion criteria were used to define the final set of papers analyzed in the next phase. Many papers were excluded from the study because they focused on EDI but did not explicitly include LGBTQIA+ students. There were also papers on EDI, but not in the scope of CS and SE education. Figure 2 summarizes the process of searching and selecting papers.

*E. Data Extraction and Synthesis*

A form was implemented to guide data extraction. In this step, the full text of each paper was accessed independently by two researchers to fill in the fields in the form described in Table 2. The process with two researchers working on the data extraction process is meant to improve the accuracy of the extraction process and, therefore, the reliability of the results. Conflicts of extraction were discussed and solved in a consensus meeting involving a third researcher when necessary.

Due to the low number of primary studies identified, no extensive method for analysis and synthesis was performed in this study. Instead, descriptive statistics and thematic analysis [27] were applied to integrate the data extracted and answer the research questions.

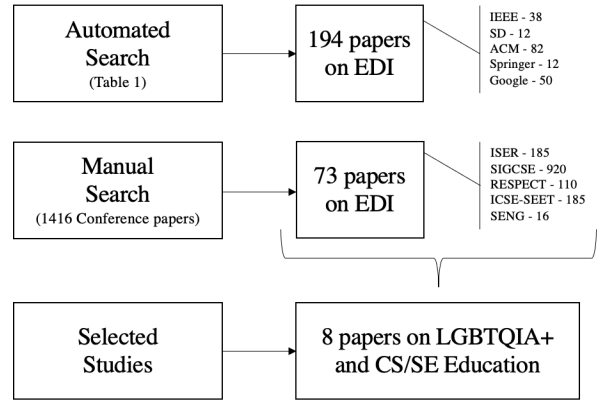

Fig. 2. Selection process.

TABLE II. DATA EXTRACTION FORM

| Data | Description |
|---|---|
| Title | Title of the paper |
| Year | Year of publication of the paper |
| Publisher | Type of publication: journal or conference |
| Country | Country of the authors |
| Study Goal | Main objective of the paper |
| Context | Undergrad or graduate |
| Experience | Main findings of the paper |
| Challenges | Challenges reported |
| Implications | Implications for academy and/or industry |
| Future | Indications of future work |

IV. RESULTS

From 267 papers focused on equity, diversity, and inclusion in CS and SE education, only eight papers presented results that explicitly referred to LGBTQIA+ students. This is a considerably low rate (only 3% of the total). Table 3 presents a summary of these papers sorted by year of publication, referred to at the end of this study following the label [LGBTQnn] to differentiate them from the reference list. Following this, each of the specific research questions is answered with the information extracted from the primary studies.

TABLE III. DEMOGRAPHICS

| Year | Study | Country | Level |
|---|---|---|---|
| 2016 | [LGBTQ01] | US | Grad and Undergrad |
| 2018 | [LGBTQ02] | US | Undergrad |
| 2018 | [LGBTQ03] | Ireland | Undergrad |
| 2018 | [LGBTQ04] | US | Undergrad |
| 2019 | [LGBTQ05] | UK | Undergrad |
| 2021 | [LGBTQ06] | US | Undergrad |
| 2021 | [LGBTQ07] | Ireland | Undergrad |
| 2022 | [LGBTQ08] | US | Undergrad |

### A. What is the evolution in the number of studies over the years?

The number of studies focused on LGBTQIA+ students in CS and SE courses is quite low. Although the number of papers focused on EDI in SE and CS education is gradually increasing, most of these studies did not (explicitly) address any aspects related to LGBTQIA+ students. In most cases, LGBTQIA+ students are cited in the study, but no significant finding or discussion is made regarding them.

By answering this research question, we demonstrated that an increase in the research on EDI in CS and SE education did not mean that research focusing on the inclusion of LGBTQIA+ students is being developed, even though this group of individuals is historically known for being excluded and discriminated against (e.g., Alan Turing). In other words, the literature addressing the challenges faced by LGBTQIA+ students in academia is insufficient and requires immediate attention. In particular, no study was identified in 2017 and 2020.

The research on SE and CS education that focus on LGBTQIA+ students is limited to three countries US (5 studies), Ireland (2 studies), and the UK (1 study). However, these studies have no apparent connection, i.e., there is no indication of collaboration among researchers or research groups. Therefore, by answering this research question, we suggest that researchers join their efforts to investigate this topic, similar to what has been observed in studies focusing on gender in CS and SE over the years.

### B. What experiences have been reported on including LGBTQIA+ students in CS and SE courses?

We identified three studies focused on increasing the sense of belonging of LGBTQIA+ students in CS and SE courses: [LGBTQ01], [LGBTQ03], and [LGBTQ07]. These studies argue that LGBTQIA+ individuals might feel unwelcome in technology courses because social aspects are rarely part of the curriculum. Sense of belonging affects self-concept and motivation to pursue a career in any field; therefore, increasing the sense of belonging among LGBTQIA+ students in CS and SE programs decreases their inclination to drop the courses. We must foster an inclusive environment where all individuals feel acceptance and belonging to attract and retain talented students in technology.

Further, two studies discussed modifications in the curriculum to promote diversity and foster inclusion among students. These studies reported experiences with creating new courses. In the first experience [LGBTQ02], the authors created a course focused on underrepresented students to provide them with a safe, inclusive environment to learn open source, designed to increase their confidence in dealing with technology. In the second experience [LGBTQ04], the authors proposed a new introductory course that explored topics designed to foster significant collaboration among students, which improved the participation and involvement of LGBTQIA+ people.

Following this, two studies reported strategies to develop modifications in the syllabus, resulting in more inclusive lectures. First, [LGBTQ06] proposed incorporating reading assignments about diversity into CS and SE courses to increase students' awareness of diversity in the software industry (e.g., the positive effects of LGBTQIA+ inclusiveness for tech companies) to inspire them in creating more inclusive environments. Second, [LGBTQ08] focused on increasing LGBTQIA+ representation by developing gender and sexual orientation-inclusive textbooks by modifying cisnormative pronouns, adapting heteronormative example scenarios, and adding historical facts about relevant LGBTQIA+ scientists in the readings.

Finally, [LGBTQ05] focused on university students' mental health and well-being and discussed interacting with data from students' smartphones to obtain insights on stress, depression, mood, suicide, and other risks. The study demonstrated that although individuals are positive towards this type of health technologies, LGBTQIA+ students in SE and CS courses are more concerned about how these technologies might track down data about their dating routines (LGBTQIA+ dating apps) and less worried about how they can obtain health support. Caring more about what people think about their sexuality than about their health demonstrates the harms of a non-supportive and non-inclusive environment for LGBTQIA+ students.

### C. What are the future challenges regarding increasing diversity and inclusion of LGBTQIA+ students in CS and SE courses?

Having only eight studies published in the past five years that focused on LGBTQIA+ students in CS and SE programs reveals the main challenge for inclusion and diversity in this context. We need more exploratory research, replication of the experiences reported in [LGBTQ01] – [LGBTQ8], and further analysis of these scenarios.

LGBTQ students feel more secure and welcomed in fields such as the social sciences and humanities [LGBTQ01]; therefore, increasing students' sense of belonging is a major challenge for CS and SE education. We need studies that compare the correlation between LGBTQ students' sense of belonging and their willingness to keep enrolled in courses across many different disciplines and use these findings to design strategies to increase inclusion [LGBTQ01] [LGBTQ03] [LGBTQ07].

As for the creation of new courses and the adaptation of learning topics to incorporate representativeness and diversity in CS and SE, the current challenges are related to determining the long-term effects of these practices, including the development of cross-sectional surveys to access the perceptions about these experiences and their effects on inclusion [LGBTQ02] [LGBTQ06] [LGBTQ08]. In addition, developing activities that go beyond the classroom, such as creating inclusive, diverse, and welcoming environments, groups, and events, is a challenge that compels immediate attention [LGBTQ04].

Finally, addressing health issues in the university environment is also a challenge; therefore, increasing diversity in CS and SE courses should be supported by discussions on the mental health of LGBTQIA+ students [LGBTQ05].

### D. What are the implications of this panorama for the software industry and research community?

The challenges described in Section IV A. indicate some of the implications of this panorama for the research community. Currently, there is a lack of studies addressing the diversity and inclusion of LGBTQIA+ people in CS and SE courses. This scenario demonstrated the need for future research on this theme. In particular, Section IV C describes challenges and directions for future research that can be used as a starting point for new investigations.

As for the industry, this panorama turns on a red light on how software development environments are now and how we expect them to be in the future. The lack of diversity in CS and SE courses is just the beginning of a pipeline propagated to software companies. Consequently, the lack of diversity in software teams can affect the technologies delivered to our diverse society, affecting clients and users on many levels.

## V. DISCUSSIONS

The arrival of Generation Z in undergraduate courses required educators to adapt and align their programs and processes with this generation's characteristics, motivations, and learning styles. Generation Z encompasses all the individuals born between 1995 and 2010 [28][29][30]. Generation Z students are defined as open-minded, determined, and thoughtful individuals motivated by how their actions affect others and learn based on exploring real-world problems [31]. Educators are often required to be aware of world events to be aligned with the perspective of these students. For SE and CS educators, it means providing students with enough innovative approaches and keeping up with technological trending.

Considering this general characteristic of the students currently enrolled in undergraduate courses, it seems counter-intuitive that fields that rely on innovation and creativity, such as computer science and software engineering, struggle with a lack of diversity and inclusion. Half of the individuals from Generation Z identify as exclusively attracted to the opposite sex [28][29], which means that Generation Z either self-identify as LGBTQIA+ individuals or are heterosexuals who tend to be more accepting of the LGBTQIA+ community. Then, why are LGBTQIA+ students feeling unwelcome in CS and SE courses when this field has Alan Turing, a gay man, frequently referred to as the *father* of *computer science*?

LGBTQIA+ inclusion in CS and SE courses is essential given the history and the contributions of LGBTQIA+ individuals to the field and due to the characteristics of Generation Z, who are the majority in undergraduate courses nowadays. Moreover, addressing the diversity crisis in software engineering should be one of the main priorities of researchers and practitioners because technology plays a crucial role in people's lives, influencing several aspects of our society, which is heterogeneous, multifaceted, and diverse [19].

Finally, diversity is essential to developing innovative ideas because diverse teams are more creative and efficacious [32]. On the other hand, inclusion supports productivity, talent retention, and teamwork [33][35]. Those are crucial elements for the software industry [34]. Therefore, if we want diverse and inclusive software development environments, we must start by fostering diversity and inclusion in CS and SE courses and propagating these characteristics from the university to the industry. The results of this mapping study demonstrate that we are behind in other areas regarding research efforts to increase LGBTQIA+ inclusion in academia.

## VI. THREATS TO VALIDITY

Bias in the study selection is one of the most common threats to validity in secondary studies [26]. We tried to reduce this bias by using a combination of both automated and manual searches. However, we acknowledge that some papers might not have been identified in our process either because they were published outside of the specific context of computer science (e.g., under STEM in general or education) or because LGBTQIA+ students were considered as being part of a broader population (e.g., gender minorities). In any of these cases, even though this might have excluded some papers from our selection process, in this study, we were targeting papers that have CS and SE LGBTQIA+ students as their primary focus and not as part of a broad population.

Moreover, it is essential to disclose that even though our search string did not use distinct terms in the acronym (i.e., lesbian, gay, transgender, etc.), studies that research any of these individuals alone are likely to use the LGBTQIA+ acronym to refer to the community. In addition, we applied the individual terms when searching for papers manually.

## VII. CONCLUSIONS

We conducted a mapping study to identify and analyze the available literature on the inclusion of LGBTQIA+ students in computer science and software engineering education. Combining automated and manual techniques, we identified 267 papers discussing EDI aspects in technology courses. However, only eight papers focused on LGBTQIA+ individuals. This lack of attention to this population, e.g., research focusing specifically and explicitly on this group of individuals, is represented in the title of this study.

In summary, the few identified studies raised concerns about how LGBTQIA+ feel unwelcomed in CS and SE courses and proposed some experiences for adapting courses and curricula to increase inclusion in academic environments. Our findings revealed the challenges of increasing diversity and inclusion in CS and SE courses. Our discussion is an invitation to reflect on why fostering inclusion in educational environments is vital to the software industry. In summary, the lack of diversity in CS and SE courses propagates in a pipeline to software companies. However, these companies depend on innovation and creativity to keep producing software products that will support society, and diversity is one key aspect of innovation.

As Generation Z is the majority in universities now, and these individuals are known for being open-minded and supportive of social problems, we trust in the success of strategies to increase diversity in CS and SE courses which will encourage the inclusion of LGBTQIA+ students in software development environments.

This study is the initial step of a broad research program that focuses on promoting equity, diversity, and inclusion in computer science and software engineering, both in academia and industrial practice. Based on these findings, our immediate future works will focus on extending this study to a) identifying experiences on including LGBTQIA+ students in CS and SE courses published as grey literature and b) identifying and comparing experiences to improve the inclusion of LGBTQIA+ students in courses from other knowledge fields, such as health and social sciences. Long-term future works will focus on proposing novel strategies and validating experiences that can be applied in the context of CS and SE courses to promote equity, diversity, and inclusion.

## ACKNOWLEDGMENT

We want to acknowledge all the LGBTQIA+ individuals pursuing a career in computer science, software engineering, and related courses. You are not alone!